\documentclass[12pt]{article}
\usepackage{amsmath}
\usepackage{amssymb}
\date{}
\begin{document}

\title{A Generalized Cole-Hopf Transformation
for Nonlinear ODES}

\author{Mayer Humi\\
Department of Mathematical Sciences,\\
Worcester Polytechnic Institute,\\
100 Institute Road,\\
Worcester, MA  0l609}

\maketitle

\begin{abstract}
We introduce a hybrid Cole-Hopf-Darboux transformation to relate solutions 
of nonlinear and linear second order differential equations and derive 
a sufficient condition for this correspondence. In particular 
we show that solutions of some nonlinear second order equations are 
related to the special functions of mathematical physics through this 
transformation. These nonlinear equations can be viewed as the 
"class of special nonlinear equations" which correspond to the linear
differential equations which define the special functions of 
mathematical physics.
%From a physical point of view this demonstrates that
%a physical system might be nonlinear even though its "states" are 
%related to solutions of a familiar linear differential equation.
\end{abstract}

\thispagestyle{empty}

\newpage

\section{Background}

In the context of partial differential equations the Hopf-Cole
transformation [1,2] and it generalizations [3,4]  has been used
extensively to linearize some nonlinear partial differential equations
such as the Burgers equation [5,6,7].

For ordinary differential equations a "similar" transformation
has been used for some time. In fact it is well known [8] that 
the Ricatti equation 
\begin{equation}
\label{1.1}
\psi(x)^{\prime}+A(x)\psi(x)^2+B(x)\psi(x)+C_1(x)=0
\end{equation}
where primes denote differentiation with respect to $x$,
can be linearized by the transformation
\begin{equation}
\label{1.2}
\psi(x)=\frac{\phi(x)^{\prime}}{A(x)\phi(x)}.
\end{equation}
The resulting linear equation for $\phi(x)$ is
\begin{equation}
\label{1.3}
A(x)\phi(x)^{\prime\prime}+(B(x)A(x)-A(x)^{\prime})\phi(x)^{\prime}+
C_1(x)A(x)^2\phi(x)=0
\end{equation}
However the operator
\begin{equation}
\label{1.4}
T=\frac{d}{dx}+A(x)\psi(x)+B(x)
\end{equation}
can be used repeatedly to generate higher order differential equations
that can be linearized by the transformation (\ref{1.2}). That is the
differential equation 
\begin{equation}
\label{1.5}
T^n\psi(x)=Q(x)
\end{equation}
where $T^n\psi(x)=T(T^{n-1})\psi(x)$ can be linearized by the 
transformation (\ref{1.2}).

In particular for $n=2$ we have
\begin{equation}
\label{1.5a}
T^2\psi=(\frac{d}{dx}+A(x)\psi(x)+B(x))(\frac{d}{dx}+A(x)\psi(x)+B(x))\psi(x).
\end{equation}
Thus for $n=2$, and $A(x)=1$, $B(x)=0$ the nonlinear differential equation 
(\ref{1.5}) becomes
\begin{equation}
\label{1.6}
\psi(x)''+3\psi(x)\psi(x)'+\psi(x)^3=Q(x).
\end{equation}
Applying the transformation (\ref{1.2}) to this equation yields
\begin{equation}
\label{1.7}
\phi(x)^{\prime\prime\prime}=Q(x)\phi(x)
\end{equation}

This shows that the transformation (\ref{1.2}) relates some nonlinear 
equations to a linear one of {\bf higher order}. However there is another 
class of transformations that are used to relate the solutions of two linear 
differential equations of the {\bf same order}. These are Darboux 
transformations [9-12] (which form the basis for the well known 
Factorization method [13]). In this case the operator that relates the 
two equations is of the form
\begin{equation}
\label{1.8}
D=C(x)+\frac{d}{dx}.
\end{equation}

Our objective in this paper is to explore the possible use of some 
"hybrid" form of (\ref{1.2}) and (\ref{1.8}) to relate the solutions of 
a nonlinear equation to those of a linear one of the {\bf same order}. 
In particular we are motivated by the fact that the solutions of 
some nonlinear second order equations are related to the special functions 
of mathematical physics through a transformation similar to (\ref{1.2}).
In a certain sense these equations define then a class of 
"{\it special nonlinear differential equations}". Furthermore we show 
that Painleve II equation has a solution that can be expressed in terms 
of Airy functions for a special set of its parameters.

%From a physical point of view this demonstrates that ambiguity might
%exist about the nature of a physical system (whether it is linear or 
%nonlinear) even if it states (or data) might be represented by functions 
%which satisfy a familiar linear differential equation.  

The plan of the paper is as follows in Sec. $2$ we present the general 
technique which relates solutions of linear and nonlinear equations. In 
Sec $3$ we specialize to a subset of this general method and provide
and intrinsic test for the applicability of the method. Sec $4$
explores the relationship between some nonlinear equations and the 
special functions of mathematical physics. Sec $5$ provides some additional 
examples of nonlinear equations whose solutions are related to those of 
a linear equation. We end up in Sec. $6$ with some conclusions.

\setcounter{equation}{0} 
\section{A Generalized Transformation}

We shall say that the solutions of the equations
\begin{equation}
\label{2.1}
\psi(x)''=S(x)+ V(x)\psi(x)+W(x)\psi(x)^2+R(x)\psi(x)^3+\lambda\psi(x)
\end{equation}
and
\begin{equation}
\label{2.2}
\phi(x)''=U(x)\phi(x)+K(x)\phi(x)'+\lambda\phi(x)
\end{equation}
are related if we can find functions $P(x)$ and $Q(x)$ so that
\begin{equation}
\label{2.3}
\psi(x)=P(x)+Q(x)\frac{\phi(x)^{\prime}}{\phi(x)}.
\end{equation}
We observe that, in principle, the terms $\lambda\psi(x)$, $\lambda\phi(x)$ 
in (\ref{2.1}), (\ref{2.2}) can be absorbed by $V(x)$ and $U(x)$
respectively. Furthermore (\ref{2.1}) can take the more general form
\begin{equation}
\label{2.1b}
\psi(x)'' = S(x)+ V(x)\psi(x)+V_1(x)\psi(x)'+W(x)\psi(x)^2+R(x)\psi(x)^3.
\end{equation}
In this case we can find $p(x)$ so that $V_1(x)=-2\frac{p(x)'}{p(x)}$.
Introducing $\xi(x)=p(x)\psi(x)$, (\ref{2.1b}) becomes
\begin{equation}
\label{2.1c}
\xi(x)''= p(x)S(x)+ (V(x)+\frac{p(x)''}{p(x)})\xi(x)+\frac{W(x)}{p(x)}\xi(x)^2
+\frac{R(x)}{p(x)^2}\xi(x)^3.
\end{equation}
which has the same form as (\ref{2.1}).

To classify those nonlinear equations (\ref{2.1}) which can be "paired"
with a linear equation of the form (\ref{2.2}) we differentiate 
(\ref{2.3}) twice and in each step replace the second order derivative 
of $\phi(x)$ by $U(x)\phi(x)+K(x)\phi(x)'+\lambda\phi(x)$. We then use 
(\ref{2.1}) to eliminate $\psi(x)''$. As a result we find that the following 
equation must hold;
\begin{equation}
\label{2.4}
a_3(x)\phi(x)^{-3}+a_2(x)\phi(x)^{-2}+a_1(x)\phi(x)^{-1}+a_0(x)=0
\end{equation}
where 
\begin{equation}
\label{2.5}
a_3(x)=-Q(x)^2R(x)+2,\,\,\,a_2(x)=-2Q(x)'- 3R(x)P(x)Q(x)^2-W(x)Q(x)^2-3K(x)Q(x)
\end{equation}
\begin{eqnarray}
\label{2.6}
a_1(x)&=&Q(x)''+2K(x)Q(x)'- \\ \notag
&&Q(x)[3P(x)^2R(x)+2U(x)+V(x)+2P(x)W(x)-K(x)'-K(x)^2+3\lambda]
\end{eqnarray}
\begin{eqnarray}
\label{2.7}
a_0(x)&=&2(U(x)+\lambda)Q(x)'+Q(x)U(x)'+P(x)''-W(x)P(x)^2-V(x)P(x)- \\ \notag
&&R(x)P(x)^3-\lambda P(x)+K(x)Q(x)(U(x)+\lambda)-S(x)
\end{eqnarray}
To satisfy (\ref{2.4}) it is sufficient to let $a_i(x)=0,\,\,i=0,1,2,3$.
We use these conditions to express $S(x),V(x),W(x),R(x),K(x)$ and $U(x)$ 
in terms of the parameters $P(x),Q(x)$. From (\ref{2.5}) we get
\begin{equation}
\label{2.8}
R(x)=\frac{2}{Q(x)^2},\,\,\,\, W(x)=-\frac{2(Q(x)'+3P(x))+3K(x)Q(x)}{Q(x)^2}.
\end{equation}
Substituting these results in (\ref{2.6}) we obtain an equation which 
we can solve for $V(x)$
\begin{eqnarray}
\label{2.9}
V(x)&=&\frac{Q(x)Q(x)''+4P(x)Q(x)'+6P(x)^2-Q(x)^2(2U(x)-3\lambda)}{Q(x)^2}+ \\ \notag
&&K(x)^2+\frac{(6P(x)+2Q(x)')K(x)}{Q(x)}+K(x)'.
\end{eqnarray}
Substituting these expressions in (\ref{2.7}) we obtain the following 
first order linear equation for $U(x)$
\begin{eqnarray}
\label{2.10}
&&Q(x)U(x)'+(2Q(x)'+2P(x)+K(x)Q(x))U(x)+P(x)''+ \\ \notag
&&2\left\{\lambda-\frac{P(x)}{Q(x)}\left(K(x)+\frac{P(x)}{Q(x)}\right)\right\}Q(x)' +2\lambda P(x)-\frac{P(x)^2}{Q(x)^2}(2P(x)+3K(x)Q(x))- \\ \notag
&&\frac{P(x)Q(x)''}{Q(x)}+K(x)(\lambda Q(x)-K(x)P(x))-P(x)K(x)'-S(x)=0.
\end{eqnarray}
Equation (\ref{2.10}) is a linear differential equation for 
$U(x)$ which can be solved by standard methods once $P(x)$ and $Q(x)$ 
and $K(x)$ have been specified.

\setcounter{equation}{0} 
\section{Solutions with $Q(x)=1$}

When $Q(x)=1$ we have $R(x)=2$ and 
\begin{equation}
\label{2.11}
V(x)=6P(x)(P(x)+K(x))-2U(x)-3\lambda+K(x)^2+K(x)',\,\,\,\, W(x)=-6P(x)-3K(x). 
\end{equation}
Eq. (\ref{2.10}) simplifies and we have 
\begin{eqnarray}
\label{2.12}
&&U(x)'+2P(x)(U(x)+\lambda-P(x)^2)+P(x)''-S(x) + \\ \notag
&&( U(x)+\lambda)K(x)-P(x)(K(x)^2+3K(x)P(x)+K(x)')=0.
\end{eqnarray}
Since this equation contains two unknown functions $U(x)$ and $K(x)$ it 
is natural to by split it into two equations
\begin{equation}
\label{2.13}
K(x)'+K(x)^2+(3P(x)-\frac{\lambda}{P(x)})K(x)+\frac{S(x)}{P(x)}=0,
\end{equation}
and
\begin{equation}
\label{2.14}
U(x)'+2P(x)(U(x)+\lambda-P(x)^2)+P(x)''+K(x)U(x) =0.
\end{equation}
Eq. (\ref{2.13}) is independent of $U(x)$ and can be solved for
$K(x)$ once $P(x)$ and $S(x)$ have been specified.
Actually (\ref{2.13}) is  a Ricatti equation whose linear second order 
form is
\begin{equation}
\label{2.15}
y(x)''+(3P(x)-\frac{\lambda}{P(x)})y(x)'+\frac{S(x)y(x)}{P(x)}=0
\end{equation}
where $K(x)=\frac{y(x)'}{y(x}$. We see that by proper choice of $P(x)$
and $S(x)$ the solution $y(x)$ can be a special functions of 
mathematical physics. In this case $K(x)$ will be the logarithmic 
derivative of such functions.

Another possible decomposition of (\ref{2.12}) is to as follows
\begin{equation}
\label{2.16}
K(x)'+K(x)^2+(3P(x)-\frac{\lambda}{P(x)})K(x)=0,
\end{equation}
and
\begin{equation}
\label{2.17}
U(x)'+2P(x)(U(x)+\lambda-P(x)^2)+P(x)''+K(x)U(x)-S(x) =0.
\end{equation}
If we then let $S(x)=-2P(x)^3$, (\ref{2.17}) becomes linear in both 
$U(x)$ and $P(x)$. As a result one may choose $P(x)$ and solve this equation
for $U(x)$ or choose $U(x)$ and solve for $P(x)$.

Another option to find solutions to (\ref{2.12}) is to cancel the nonlinear
terms in $P(x)$ by making the ansatz 
\begin{equation}
\label{3.1}
S(x)=-3P(x)^2 K(x)-2P(x)^3
\end{equation}
the equation then becomes 
\begin{equation}
\label{3.2}
P(x)''+(2U(x)+2\lambda-K(x)'-K(x)^2)P(x)+K(x)(U(x)+\lambda)+U(x)' =0.
\end{equation}
which is a linear equation for $P(x)$ once $U(x)$ and $K(x)$ were chosen.

In the following we use all these strategies to to find nonlinear 
differential equations whose solutions are related to those of a linear 
differential equation by the transformation (\ref{2.3}). 

We now consider the following practical question: Suppose one is 
considering a nonlinear differential equation of the form given by 
(\ref{2.1}). Under what conditions one can find a linear differential 
equation (\ref{2.2}) whose solutions are related to it  by (\ref{2.3})
with $Q(x)=1$. 

{\bf Theorem}: A sufficient condition for the solution of (\ref{2.1}) to be 
related to an equation of the form (\ref{2.2}) by (\ref{2.3}) with 
$Q(x)=1$ is that $R(x)=2$ and 
\begin{equation}
\label{4.1}
S(x)=\frac{1}{2}\left\{-V(x)'+\frac{1}{3}\left[-W(x)''+(V(x)+W(x)'+\lambda)W(x)\right]-\frac{1}{27}W(x)^3\right\}
\end{equation}
When this condition is satisfied one can choose $K(x)=0$ and
\begin{equation}
\label{4.2}
U(x)=-\frac{1}{2}V(x)+\frac{1}{12}W(x)^2-\frac{3}{2}\lambda
\end{equation}

We observe that the condition (\ref{4.1}) is an intrinsic condition on 
the coefficients of (\ref{2.1}).

% it is NOT a necessary condition since (2.11) (2.12) provide only 
%a sufficient conditon for the solution of (2.6)

{\bf Proof}: From (\ref{2.11}) we have 
\begin{equation}
\label{4.3}
P(x)=-\frac{1}{6}W(x)-\frac{1}{2}K(x)
\end{equation}
Substituting this expression in the formula for $V(x)$ it follows that
\begin{equation}
\label{4.4}
V(x)-\frac{1}{6}W(x)^2+\frac{1}{2}K(x)^2+2U(x)+3\lambda-K(x)'=0
\end{equation}
Solving (\ref{4.4}) for U(x) we find that
\begin{equation}
\label{4.5}
U(x)=-\frac{1}{2}V(x)+\frac{1}{12}W(x)^2-\frac{1}{4}K(x)^2-
\frac{3}{2}\lambda+\frac{1}{2}K(x)'
\end{equation}
Finally substituting these expressions for $P(x)$ and $U(x)$ in 
(\ref{2.12}) the condition (\ref{4.1}) follows. If we let $K(x)=0$ in
(\ref{4.3}),(\ref{4.4}) and (\ref{4.5}) the expression for $U(x)$
reduces to the one given by (\ref{4.2}). $\blacksquare$.

%When $Q(x)\ne 1$, $R(x)=\frac{2}{Q(x)}$. The following  
%constraint replaces (\ref{4.1}) and can be derived by the same 
%procedure described above.
%\begin{eqnarray}
%\label{4.6}
%S(x)&=& -\frac{1}{54}Q(x)^4W(x)^3+\frac{1}{6}Q(x)^3W(x)W(x)'+ \\ \notag
%&&\left[\frac{5}{18}Q(x)'W(x)^2+\frac{1}{6}V(x)W(x)-\frac{1}{6}W(x)''+\frac{1}{6}\lambda W(x)\right]Q(x)^2 \\ \notag
%&&\left[-\frac{1}{3}W(x)Q(x)''-\frac{2}{3}W(x)'Q(x)'-\frac{1}{2}V(x)'\right]Q(x)-\frac{1}{3}W(x)(Q(x)')^2-\frac{2\lambda}{3}Q(x)'- \\ \notag
%&&\frac{2}{3}Q(x)'V(x)+\frac{1}{6}Q(x)'''-\frac{Q(x)'Q(x)''}{6Q(x)}+
%\frac{2(Q(x)')^3}{27Q(x)^2}
%\end{eqnarray}

\setcounter{equation}{0} 
\section{Relationships to the Special Functions}

In this section we explore the relationship between some nonlinear 
equations and the function of mathematical physics. The purpose of
our treatment is to highlight this new relationship and is not
comprehensive. In all cases we let $Q(x)=1$.

{\bf Case 1: Exponential and Trigonometric Functions}:

We start by considering the harmonic oscillator equation,
\begin{equation}
\label{3.17a}
\phi(x)''=-\omega^2\phi(x), \,\,\,\, \omega \ne 0.
\end{equation}
For this equation $U(x)=-\omega^2$, $K(x)=0$, and $\lambda=0$. 
Hence from (\ref{3.2})
we have
$$
P(x)''-2\omega^2 P(x)=0.
$$
Therefore
$$
P(x)=C_1e^{\sqrt{2}\omega x}+C_2e^{-\sqrt{2}\omega x}
$$
Computing $V(x)$ $W(x)$ and $S(x)$ using (\ref{2.11}) and (\ref{3.1})
we finally obtain the following differential equation for $\psi(x)$
\begin{equation}
\label{3.18a}
\psi(x)''-(6P(x)^2+2\omega^2)\psi+6P(x)\psi(x)^2-2\psi(x)^3=2P(x)^3
\end{equation}
whose solutions of are related to those of (\ref{3.18a}) by (\ref{2.3}).

If we consider
\begin{equation}
\label{3.19a}
\phi(x)''=\omega^2 \phi(x)=0,\,\,\,\, \omega \ne 0,
\end{equation}
then
$$
P(x)=C_1\sin(\sqrt{2}\omega x)+C_2\cos(\sqrt{2}\omega x)
$$
and the differential equation for $\psi(x)$ becomes
\begin{equation}
\label{3.20a}
\psi(x)''-(6P(x)^2-2\omega^2)\psi(x)+6P(x)\psi(x)^2-2\psi(x)^3=-2P(x)^3
\end{equation}
For the case where $\omega=0$ i.e.
\begin{equation}
\label{3.19b}
y(x)''=0
\end{equation}
then 
$$
P(x)=C_1+C_2x.
$$
and the equation for $\psi(x)$ is
\begin{equation}
\label{3.20b}
\psi(x)''-6P(x)^2\psi(x)+6P(x)\psi(x)^2-2\psi(x)^3=-2P(x)^3
\end{equation}

Similar treatment can be made for the general second order equation 
with constant coefficients. We omit the details.

%For the more general case of a second order equation with constant 
%coefficients
%\begin{equation}
%\label{3.20}
%y(x)''=by(x)'+a*y(x), \,\,\,\, \omega^2=2a-b^2 > 0
%\end{equation}
%we find
%$$
%P(x)=C_1\cos\omega x+C_2\sin\omega x-\frac{ab}{\omega^2},\,\,\,\,
%V(x)=6P(x)^2+6bP(x)
%$$
%$$
%W(x)=-6P(x)-3b,\,\,\,\ S(x)=-P(x)^2(3b-2P(x))
%$$
%%The other cases can be treated similarly.

{\bf Case 2: Legendre polynomials}

The differential equation for the Legendre polynomials is
\begin{equation}
\label{3.3}
\phi(x)''=\frac{2x}{1-x^2}\phi(x)'-\frac{n(n+1)}{1-x^2}\phi(x).
\end{equation}
That is $U(x)=-\frac{n(n+1)}{1-x^2}$, $K(x)=\frac{2x}{1-x^2}$ and
$\lambda=0$. Substituting this in (\ref{3.2}) we find that a particular 
solution for $P(x)$ is
\begin{equation}
\label{3.4}
P(x)=-\frac{2n(n+1)x}{(n^2+n-2)(1-x^2)},\,\,\,\, n\ne 1
\end{equation}
(The solution for $n=1$ is available but we  shall not elaborate on 
it further). Using (\ref{2.11}) we have
\begin{equation}
\label{3.5}
V(x)=-\frac{48n(n+1)x^2}{(n+2)^2(n-1)^2(1-x^2)^2}+
\frac{2n(n+1)}{1-x^2}+\frac{2(2x^2+1)}{(1-x^2)^2}.
\end{equation}
From (\ref{3.1}) and (\ref{2.11}) we obtain
\begin{equation}
\label{3.6}
S(x)=\frac{8n^2(n+1)^2(n+3)(n-2)x^3}{(n+2)^3(n-1)^3(x^2-1)^3}.
\end{equation}
\begin{equation}
\label{3.7}
W(x)=-6P(x)-3K(x).
\end{equation}
For these functions $S(x),\,V(x),\,W(x)$ (and $R(x)=2$) the solutions of 
(\ref{2.1}) are related to the solutions of Legendre equation by 
the transformation (\ref{2.3}).

{\bf Case 3: Bessel Functions}

The differential equation for Bessel functions is
\begin{equation}
\label{3.8}
\phi(x)''=-\frac{\phi(x)'}{x}-\frac{x^2-p^2}{x^2}\phi(x)
\end{equation}
That is $U(x)=-\frac{x^2-p^2}{x^2}$, $K(x)=-\frac{1}{x}$ and $\lambda=0$. 
Substituting  these in (\ref{3.2}) we find that the general solutions for 
$P(x)$ with $p=0,1$ respectively are 
\begin{equation}
\label{3.9}
P_0(x)=\frac{C_1e^z(\sqrt{2}-2x)+C_2e^{-z}(2x+\sqrt{2})}{x}+
\frac{1}{2x},\,\,\,\, z=\sqrt{2}x,
\end{equation}
\begin{equation}
\label{3.10}
P_1(x)=C_1e^z+C_2e^{-z}-
\frac{ze^z\Gamma(0,z)-ze^{-z}\Gamma(0,-z)}{2x}+\frac{3}{2x},
\end{equation}
where
$$
\Gamma(0,z)=\displaystyle\int_z^{\infty} t^{-1}e^{-t} dt
$$
If we choose for $n=0$ the special solution $P_0(x)=\frac{1}{2x}$,
we find that
$$
V(x)=2+\frac{1}{2x^2},\,\,\,\, W(x)=0,\,\,\, S(x)=\frac{1}{2x^3}.
$$
The differential equation for $\psi(x)$ is
\begin{equation}
\label{3.10a}
\psi(x)''=\frac{1}{2x^3}+(2+\frac{1}{2x^2})\psi+2\psi(x)^3.
\end{equation}
The solutions of this nonlinear equation are related to the Bessel functions
of order zero by the transformation (\ref{2.3}).

{\bf Case 4: Hermite polynomials}

The differential equation for Hermite polynomials is
\begin{equation}
\label{3.11}
\phi(x)''=2x\phi(x)'-2n\phi(x).
\end{equation}
Hence $U(x)=-2n$, $K(x)=2x$. Substituting these in (\ref{3.2}) we find that
an explicit particular (and general) solution for $P(x)$ in terms of 
elementary functions can be obtained for even $n$. For $n=2,4$ these 
particular solutions are
\begin{equation}
\label{3.12}
P_2(x)=-\frac{8\left[\pi(1/4+x^2)\exp(x^2)erf(x)+\sqrt{\pi} x\right]}{\sqrt{\pi}},
\end{equation}
\begin{equation}
\label{3.14}
P_4(x)=-\frac{64\left[\pi(3/16+3/2x^2+x^4)\exp(x^2)erf(x)+\sqrt{\pi} x(1+x^2)\right]}{3\sqrt{\pi}},
\end{equation}
where 
$$
erf(x)=2\frac{\displaystyle\int_0^x \exp(-t^2) dt}{\sqrt{\pi}}.
$$
The computation of the functions $S(x),\,V(x),\,W(x)$ 
using (\ref{2.11}) and (\ref{3.1}) is straightforward.

{\bf Case 5: Laguerre Polynomials}:

The differential equation for the Laguerre polynomials is
\begin{equation}
\label{3.15}
\phi(x)''=-\frac{1-x}{x}\phi(x)'-\frac{n}{x}\phi(x).
\end{equation}
That is $U(x)=-\frac{n}{x}$, $K(x)=-\frac{1-x}{x}$ and $\lambda=0$.
Substituting  these in (\ref{3.2}) we find that a particular solution for
$P(x)$ with $n=2$ is
\begin{equation}
\label{3.16}
P_2(x)=-{\Gamma(0,-x)e^{-x}}{x}+
e^x\Gamma(0,x)(\frac{1}{x}+2x-2)-2+\frac{2}{x}
\end{equation}
We omit the  computation of $V(x)$, $W(x)$ and $S(x)$ using (\ref{2.11}) 
and (\ref{3.1}) which is straightforward. 

{\bf Case 6: Painleve II Equation}

Painleve II equation is
\begin{equation}
\label{3.17}
\psi(x)''=2\psi(x)^3+x\psi(x)+a
\end{equation}
where $a$ is a constant. In the notation of the previous sections
$V(x)=x$, $W(x)=0$, $R(x)=2$, $S(x)=a$ and $\lambda=0$. With $Q(x)=1$
(\ref{3.16}) satisfies the constraint (\ref{4.1}) when $a=-\frac{1}{2}$.

With $K(x)=0$, (\ref{4.2}) and (\ref{4.3}) yield under the 
present settings $P(x)=0$ and
\begin{equation}
\label{3.18}
U(x)=-\frac{1}{2}V(x)=-\frac{x}{2}
\end{equation}
i.e the differential equation for $\phi(x)$ is
\begin{equation}
\label{3.19}
\phi(x)''=-\frac{x}{2}\phi(x)
\end{equation}
whose general solution is
\begin{equation}
\label{3.20}
\phi(x)=C_1Ai(-\frac{x}{2^{1/3}})+C_2Bi(-\frac{x}{2^{1/3}})
\end{equation}
where $Ai(x),\,Bi(x)$ are the Airy wave functions.
This solution is related to the solution of (\ref{3.17}) by the
transformation (\ref{2.3}) (with $P(x)=0$). We observe that other solutions 
of this equation are possible if we let $K(x)\ne 0$

\setcounter{equation}{0} 
\section{Some Special Cases}

In this section we present some explicit solutions to the equations which 
pair the solution of a nonlinear equation with a linear equation using
the algorithm which was presented in the previous sections. In all cases we
let $Q(x)=1$.

{\bf Example 1}: In this example we let $K(x)=0$ and $S(x)=0$
and use (\ref{2.12}). Choosing $P(x)=\frac{b}{x^n}$ we obtain for $n=1$
\begin{equation}
\label{2.28}
U(x)=C_1x^{-2b}+\frac{b(b+1)}{x^2}-\lambda ,
\end{equation}
and the corresponding expression for $V(x)$ is
$$
V(x)=2C_1x^{-2b}+\frac{2b(2b-1)}{x^2}-\lambda.
$$
For $n \ne 1$ we have
\begin{equation}
\label{2.29}
U(x)=C_1\exp\left(\frac{2bx^{1-n}}{n-1}\right)+\frac{nb}{x^{1+n}}+
\frac{b^2}{x^{2n}}-\lambda,
\end{equation}
and 
$$
V(x)=6\left(\frac{b}{x}\right)^{2n}-2C_1\exp\left(\frac{2bx^{1-n}}{n-1}\right)
-\frac{2bn}{x^{1+n}}-\frac{2b^2}{x^{2n}}-\lambda.
$$

For $n=1$ the equation for $\phi(x)$ with $C_1=0$ is
\begin{equation}
\label{2.30}
\phi(x)''-\frac{b(b+1)}{x^2}\phi(x) = 0,
\end{equation}
whose general solution is
\begin{equation}
\label{2.31}
\phi(x)=D_1x^{b+1}+D_2x^{-b}.
\end{equation}
The corresponding nonlinear differential equation for $\psi(x)$ is
\begin{equation}
\label{2.32}
\psi(x)''-\frac{2b(2b+1)}{x^2}\psi(x) +\frac{6b\psi(x)^2}{x}-2\psi(x)^3 = 0.
\end{equation}
In this case the explicit relationship between $\phi(x)$ and $\psi(x)$
(as postulated in (\ref{2.3})) is
\begin{equation}
\label{2.33}
\psi(x)= \frac{b}{x}+\frac{\phi(x)'}{\phi(x)}.
\end{equation}
It is straightforward to verify that this is actually a solution of 
(\ref{2.32}).

{\bf Example 2}: Let $P(x)=\frac{-2x}{3}$, $S(x)=\frac{-4nx}{3}$ and 
$\lambda=0$. Eq. (\ref{2.15}) becomes
\begin{equation}
\label{2.18}
y(x)''-2xy(x)'+2ny(x)=0
\end{equation}
which is the differential equation for the Hermite polynomials. Hence
\begin{equation}
\label{2.19}
K(x)=\frac{H_n(x)'}{H_n(x)},
\end{equation}
where $H_n(x)$ is Hermite polynomial of order $n$. Substituting these 
results in (\ref{2.14}) leads to the following general solutions for $U(x)$
\begin{equation}
\label{2.20}
U_n(x)= \frac{\left[C_1-\int\frac{16}{27}x^3H_n(x)\exp(-\frac{2x^2}{3}) dx\right]
\exp(\frac{2x^2}{3})}{H_n(x)}.
\end{equation}
The integral in (\ref{2.20}) can be computed explicitly for different
values of $n$. For $H_0(x)=1$ we have 
$$
U_0(x)=C_1\exp(\frac{2x^2}{3})-\frac{3}{8}(3+2x^2)
$$
For $H_2(x)=4x^2-2$
$$
U_2(x)=\frac{C_1\exp(\frac{2x^2}{3})-\frac{3}{4}(4x^4+10x^2+15)}{4x^2-2}
$$
etc.

From these expressions for $K(x)$ and $U(x)$ we find that
\begin{equation}
\label{2.21}
V_n(x)=\frac{8x^2}{3}-2U_n(x)+\frac{H_n(x)''-4xH_n(x)'}{H_n(x)},\,\,\,\,
W_n(x)=4x-\frac{3H_n(x)'}{H_n(x)}
\end{equation} 
We conclude that the solutions of the nonlinear equation (\ref{2.1})
with $R(x)=2$ and $V_n(x),\,W_n(x)$ in (\ref{2.21}) are related to those of 
the linear differential equation (\ref{2.2}) with $U(x)$ and $K(x)$
given by (\ref{2.20}) and (\ref{2.19}) by the transformation (\ref{2.3}).

{\bf Example 3}: In this example we let $K(x)=0$, $P(x)=a+bx$,
$S(x)=-2P(x)^3$ and use (\ref{2.17}) to compute $U(x)$. This yields
\begin{equation}
\label{2.22}
U(x)=C_1\exp[-(2a+bx)x]-\lambda,\,\,\,V(x)=6(a+bx)^2-\lambda-2C_1
\exp[-(2a+bx)x])
\end{equation} 
and $W(x)=-6P(x)$ (from (\ref{2.8})).

{\bf Example 4}: Here as in the previous example we let $K(x)=0$,
$S(x)=-2P(x)^3$ and $\lambda=0$. However we now let $U(x)=-\frac{a}{x^2}$ 
and use (\ref{2.17}) to compute $P(x)$. For $a \ne 1$ we obtain
\begin{equation}
\label{2.23}
P(x)=\frac{a}{(a-1)x}+C_1x^{1/2+\beta}+C_2x^{1/2-\beta}\,\,\,\,
\beta=\sqrt{1+8a}.
\end{equation} 
For $a=1$
\begin{equation}
\label{2.24}
P(x)=\frac{6\ln x+2}{x}+C_1x^2+\frac{C_2}{x}.
\end{equation} 
The corresponding $V(x)$ in this case (viz. $a=1$) is
\begin{equation}
\label{2.25}
V(x)=\frac{2}{x^2}+P(x)^2.
\end{equation} 

If we reverse the roles of $P(x)$ and $U(x)$ i.e fix $P(x)=-\frac{a}{x^2}$
and use (\ref{2.17}) to compute $U(x)$ we find
\begin{equation}
\label{2.26}
U(x)=C_1\exp(-\frac{2a}{x})-3\left(\frac{1}{x^2}-\frac{1}{ax}+\frac{1}{2a^2}\right).
\end{equation} 

{\bf Example 5}: Under the same settings of example $9$ we specify
$U(x)=2\exp(2ax)$ and use (\ref{2.17}) to compute $P(x)$. We find
\begin{equation}
\label{2.27}
P(x)=C_1J_0(z)+C_2Y_0(z),\,\,\,\, z=\frac{2\exp(ax)}{a}
\end{equation} 
where $J_0$ and $Y_0$ are Bessel functions of order zero of the first 
and second kind. The corresponding values of $V(x)$ and $W(x)$ are
$$
V(x)=6P(x)^2-4e^{2ax},\,\,\,\, W(x)=-6P(x).
$$

\section{Conclusions}

We demonstrated in this paper that the hybrid "Cole-Hopf-Darboux" operator
(\ref{2.3}) can be used to relate the solutions of some nonlinear and linear
second order differential equations. The algorithm is straightforward to
apply and we presented several examples which demonstrated the scope
of the method and its potential interest for problems in mathematical physics. 
From a perspective those nonlinear differential equation whose 
solutions can be expressed in terms of the special functions of 
mathematical physics (through the transformation (\ref{2.3})
should obviously be viewed as a new "special class of nonlinear equations".

Research into the possible application of the transformation (\ref{2.3})
to partial differential equations is on going.

\section*{References}

\begin{itemize}
\item[1] E. Hopf - The partial differential equation $u_t+uu_x=u_{xx}$,
Commun. Pure Appl. Math {\bf 3}, pp.201-230 (1950)

\item[2] J.D. Cole, On a quasi-linear parabolic equation occurring in 
aerodynamics, Quart. Appl. Math. {\bf 9}, pp.225-236 (1951)

\item[3] P. L. Sachdev-A generalised Cole-Hopf transformation for nonlinear 
parabolic and hyperbolic equations,
ZAMP {\bf 29}, No 6 (1978), pp. 963-970, DOI: 10.1007/BF01590817

\item[4] B. Mayil Vaganan- Cole-Hopf Transformations for Higher Dimensional 
Burgers Equations With Variable Coefficients, Studies in Applied Mathematics,
DOI: 10.1111/j.1467-9590.2012.00551.x (in press) (2012)

\item[5] G Kaniadakis and A M Scarfone - Cole-Hopf-like transformation for 
Schrodinger equations containing complex nonlinearities,
J. Phys. A: Math. Gen. 35, p. 1943 (2002).

\item[6] B. Gaffet-On the integration of the self-similar equations and
the meaning of the Cole-Hopf Transformation J. Math. Phys. 27, 2461 (1986)

\item[7] Wen-Xiu Ma- An exact solution to two-dimensional 
Korteweg-de Vries-Burgers equation,  J. Phys. A: Math. Gen. 26 L17 (1993).

\item[8] E.L. Ince - Ordinary Differential Equations, Dover Publications, 
New-York, 1956

\item[9] C. Gu, H. Chaohao and Z. Zhou - Darboux Transformations in
Integrable Systems, Springer, New-York (2005)

\item[10] M. Humi - Separation of coupled systems of differential equations by
Darboux transformation.  J. of Physics A  {\bf 18}, p. 1085 (1985).

\item[11] M. Humi - Darboux transformations for Schroedinger equations
in two variables, J. Math. Phys. {\bf 46}, 083515 (2005) (8 pages).

\item[12] M. Humi- Separation of Coupled Systems of Schroedinger equations
by Darboux transformation, Reviews in Mathematical Physics
Vol. 24, No. 3, 14 pages (2012)

\item[13] L. Infeld and T.E. Hull - The Factorization method,
Rev. Mod. Phys. {\bf 23}, p.21 (1951).

\end{itemize}

\end{document}